	\documentclass[twocolumn,showpacs,superscriptaddress,prb]{revtex4-1}

\usepackage[dvips]{graphicx}	% Include figure files
\usepackage{dcolumn}			% Align table columns on decimal point
\usepackage{bm}					% bold math
\usepackage{epsfig}				%Use .eps figures
\usepackage{color}

\begin{document}
%\title{High Field SANS measurements of the Vortex Lattice in YBa$_{2}$Cu$_{3}$O$_{7}$}
%\title{Temperature and field driven changes in the Vortex Lattice structure in high field studies on YBa$_{2}$Cu$_{3}$O$_{7}$}
\title{High magnetic field studies of the Vortex Lattice structure in YBa$_{2}$Cu$_{3}$O$_{7}$}

\author{A.S.~Cameron}
\affiliation{School of Physics and Astronomy, University of Birmingham, Edgbaston, Birmingham, B15 2TT, UK.}

\author{J.S.~White}
\affiliation{Laboratory for Neutron Scattering and Imaging, Paul Scherrer Institut, CH 5232 Villigen, Switzerland.}

\author{A.T.~Holmes}
%\affiliation{School of Physics and Astronomy, University of Birmingham, Edgbaston, Birmingham, B15 2TT, UK.}
\author{E.~Blackburn}
%\affiliation{School of Physics and Astronomy, University of Birmingham, Edgbaston, Birmingham, B15 2TT, UK.}
\author{E.M.~Forgan}
%\affiliation{School of Physics and Astronomy, University of Birmingham, Edgbaston, Birmingham, B15 2TT, UK.}
\author{R. Riyat}
\affiliation{School of Physics and Astronomy, University of Birmingham, Edgbaston, Birmingham, B15 2TT, UK.}

\author{T.~Loew}
\affiliation{Max Planck Institut f\"{u}r Festk\"{o}rperforschung, D-70569 Stuttgart, Germany.}
\author{C.D.~Dewhurst}
\affiliation{Institut Laue-Langevin, 6 rue Jules Horowitz, 38042 Grenoble, France.}
\author{A.~Erb}
\affiliation{Walther Meissner Institut, BAdW, D-85748 Garching, Germany.}

\begin{abstract}
We report on small angle neutron scattering measurements of the vortex lattice (VL) in twin-free YBa$_{2}$Cu$_{3}$O$_{7}$, extending the previously investigated maximum magnetic field of \textless11~T up to 16.7~T with the field applied parallel to the \textbf{c} axis. This is the first microscopic study of vortex matter in this region of the superconducting phase. We find the high field VL displays a rhombic structure, with a field-dependent coordination that passes \emph{through} a square configuration with no lock-in to a field-independent structure. The VL pinning reduces with increasing temperature, but is seen to affect the VL correlation length even above the irreversibility temperature of the lattice structure. At high field and temperature we observe a VL melting transition, which appears to be first order. The vortex liquid phase above the transition does not give a detectable SANS signal.
\end{abstract}


% PACS, the Physics and Astronomy Classification Scheme.
%Use showkeys class option if keyword display desired

\date{\today}

\maketitle

\section{Introduction}

\noindent 
% Modified from White etal PRB
YBa$_{2}$Cu$_{3}$O$_{7-\delta}$ (YBCO) was the first high-$T_c$ superconductor in which the mixed state was studied by small-angle neutron scattering (SANS)~\cite{For90}. In the more than 25 years since the discovery of YBCO, sample quality has greatly improved and the available field range has increased, allowing increasingly sophisticated studies of the mixed state by a variety of techniques~\cite{Gam87,Mag95,Fis97,Son97,Son99,Aus08,For90,Yet93a,Yet93b,Kei93,Kei94,For95,Aeg98,Joh99,Bro04,Sim04,Whi08,Whi09,Whi11}. In particular, SANS measurements have given direct microscopic information about  vortex lines in the mixed state in bulk YBCO~\cite{For90,Yet93a,Yet93b,Kei93,Kei94,For95,Aeg98,Joh99,Bro04,Sim04,Whi08,Whi09,Whi11}. These measurements lead to understanding of the physics of the vortex lattice (VL) and the effects of crystal anisotropy, thermal excitation and pinning on the VL structure and its thermal melting. Furthermore, the observed properties of the VL can give information about the superconducting pairing state in this material and its evolution with field.  Here we report on SANS studies of the VL in fully-oxygenated YBa$_{2}$Cu$_{3}$O$_{7}$ (YBCO$_{7}$). This material is slightly over-doped, and pinning on oxygen vacancies of magnetic vortex lines in the mixed state is greatly reduced by full oxygenation. We extend the field range of previous SANS studies from 10.8~T to 16.7~T, providing the first microscopic structural information about the behavior of the VL at such fields in a high-$T_c$ material.

YBCO has an orthorhombic crystal structure containing two-dimensional CuO$_{2}$ $ab$-plane layers, and one-dimensional CuO chain layers  running along the \textbf{b} direction. These are stacked along the \textbf{c}-direction, separated by Y and BaO layers. As in all high-$T_c$ materials, superconductivity with a nodal gap structure is mainly localized in the  CuO$_{2}$ layers. However, for optimally- and over-doped compositions, the CuO chains display both long-range order along the crystal \textbf{b}-axis, and metallic behavior.~\cite{Bas05} As a consequence, it is expected that the electronic states primarily associated with the chains are also superconducting below $T_{c}$. This picture is supported by reported values of the in-plane penetration depth ratio, $\gamma_{\lambda}~(=\lambda_{a}/\lambda_{b}\propto\sqrt{m_{a}^{\ast}/m_{b}^{\ast}})$,  which lie in the range 1.2-1.6.~\cite{Bas95,Gag97,Age00,Bro04,Aus08,Whi09,Whi11} The \emph{direction} of this anisotropy is that expected if there is a contribution to the superfluid density along  \textbf{b} from the chain carriers. The observed anisotropy cannot be due the plane carriers, because the anisotropy between the \textbf{a} and \textbf{b} hopping matrix elements for these carriers given by LDA band calculations~\cite{Lu01} is close to unity and in the wrong direction.  It has been suggested that the chain pairing may arise from a proximity-effect single-particle coupling between chain and plane states.~\cite{Atk95,Atk08,Kon10}, but an alternative cause is a Josephson pair-coupling~\cite{Xiang96, Zhang94}. It has been pointed out~\cite{Whi14} that the superfluid density along \textbf{a} and \textbf{b} both show a $d$-wave temperature-dependence at low $T$~\cite{Whi09,Whi11}. This is consistent with intrinsically superconducting chains coupled to the planes by a Josephson-type pair tunneling.~\cite{Xiang96} In this case, the $d$-wave behavior for the \textbf{b}-direction 
implies that there are nodes in the superconducting gap for the chain carriers also. The positions of these nodes on the chain Fermi surface would not be fixed by symmetry and could move as a function of field and temperature. 
A further consequence of the crystal orthorhombicity is that the predominantly $d_{x^{2}-y^{2}}$ order-parameter of the plane carriers \emph{must} contain a finite additional $s$-wave component.~\cite{Tsu00} This implies that the nodes on the plane Fermi surface are not exactly at $45^{\circ}$ to the crystal axes. Evidence for this admixture is provided by phase sensitive,~\cite{Kir06} tunneling,~\cite{Smi05} and $\mu$SR~\cite{Kha07} studies.

As-grown YBCO crystals are crystallographically twinned; the twin planes running at approximately $45^{\circ}$ between adjacent \textbf{a}- and \textbf{b}-domains act as pinning centers. These have had a clear influence on the results from almost all previous SANS studies~\cite{For90,Yet93a,Yet93b,Kei93,Kei94,For95,Aeg98,Joh99,Bro04,Sim04,Whi08}. de-twinned samples have only been used in the most recent investigations ~\cite{Whi09,Whi11}. These show \emph{first-order} phase transitions between different VL phases in contrast to the continuous transitions reported in previous studies~\cite{Bro04,Whi08}. Our present samples are de-twinned, and allow us to follow the \emph{intrinsic} behavior of the VL in YBCO into a new field range, where the vortex cores are closer together and the effect of the order parameter anisotropy on the VL structure is expected to be more pronounced.  In addition to this,  the \emph{intensity} of the SANS signal as a function of field and temperature depends on magnitude of the spatial variation of magnetic field in the mixed state. This is measured in terms of the VL ``form factor", which reflects the coherence length, $\xi$, of the Cooper pairs, the magnetic penetration depth $\lambda$, and the departure of the vortex lines from straightness, represented by a Debye-Waller factor. For our field range it is expected that the VL undergoes a melting transition to a vortex liquid on heating. SANS data can complement existing thermodynamic measurements~\cite{Rou98}, by adding a unique insight into the VL structure and perfection near the melting transition.

\section{Experimental}

%\subsection*{Sample Preparation \& Characterisation}
The sample was a mosaic of six single crystals of de-twinned YBa$_{2}$Cu$_{3}$O$_{7}$ with a total mass of $\sim$20~mg. Each crystal was grown from a molten flux of BaCO$_{3}$, CuO and Y$_{2}$O$_{3}$ in BaZrO$_{3}$ crucibles \cite{Erb96}. Upon cooling from the high temperature tetragonal phase to the orthorhombic phase, the inequality between the crystal \textbf{a} and \textbf{b} axes results in crystallographic twinning. It has been seen in previous SANS studies \cite{Bro04} that vortex lines are strongly pinned to the twin planes, dominating the VL orientation at the expense of other physical phenomena. To remove this effect, these crystals were de-twinned through the application of uniaxial stress at $500^{\circ}$ C for $24$ hours \cite{Lin92, Hin07}. The crystals were then oxygenated under an O$_{2}$ atmosphere of $100$ bar at $300^{\circ}$ C for $150$ hours \cite{Erb99}. A crystal from the mosaic gave a zero field $T_{c}$ of $89.0$ K by SQUID magnetometry in a field of 1~mT, with a 90\% superconducting transition width of 2~K. The mosaic was mounted on a $1$ mm thick pure aluminum plate, with the crystal \textbf{c} axis perpendicular to the plate and the \textbf{a} direction co-aligned between crystals.

%\subsection*{Experimental Details}
The experimental work was carried out on the D22 instrument at the Institut Laue-Langevin in Grenoble, France. The mosaic was mounted on a variable-temperature stage in a horizontal-field cryomagnet of novel design~\cite{Hol12}, with the crystal \textbf{c} axes parallel to the applied field and the \textbf{a} axes vertical. SANS measurements were performed with the neutron beam entering at a range of angles close to the field direction. Neutrons of mean wavelength of 6-10 \AA$\,$ were used with a 10\% full width at half maximum (FWHM) wavelength spread, collimated over a distance of 8~m.  

For all measurements, the vortex lattice was prepared using the oscillation field cooled (OFC) procedure, where the sample is cooled through $T_c$ in an applied field with a small ($\sim$0.1~\%) oscillation about the required field. This method was chosen because a previous investigation \cite{Whi09} showed that this allowed the VL to find its preferred structural free energy minimum for an effective temperature lower than for cooling in a static field. The diffraction patterns were acquired by rocking the sample and field together through the required Bragg conditions, producing what is known as a `rocking curve', shown in the inset of Fig. 2. Background measurements were taken above $T_{c}$ and subtracted from the in-field measurements to leave only the signal from the vortex lattice.  

Data visualization and analysis were performed using the GRASP analysis package~\citep{Dew03}.  This allowed the determination of the fitted positions and angles of VL diffraction spots, and the evaluation of rocking curve widths and intensities integrated over rocking curves. Within the mixed state, the local field is expressed as a sum over spatial Fourier components at the  two-dimensional reciprocal lattice vectors \textbf{q} of the VL. The magnitude of a Fourier component $F(\textbf{q})$ is known as the form factor at wavevector \textbf{q}. Its value is obtained from the integrated intensity, $I_{\textbf{q}}$, of a VL reflection as the VL is rocked through the Bragg  condition for a diffracted spot at scattering vector \textbf{q}. The vortex lattice form factor, $F(\textbf{q})$ is related to the integrated intensity, $I(\textbf{q})$, via the relationship~\cite{Chr77}:
\begin{equation}
I_{\textbf{q}} = 2\pi V\phi \big(\frac{\gamma}{4} \big)^{2} \frac{\lambda_{n}^{2}}{\Phi_{0}^{2}q} |F(\textbf{q})|^{2}.
\label{FF}
\end{equation}
Here, $V$ is the illuminated sample volume, $\phi$ is the incident neutron flux, $\lambda_{n}$ is the neutron wavelength, $\gamma$ is the magnetic moment of the neutron in nuclear magnetons ($=1.91$), and $\Phi_{0} = h/2e$ is the flux quantum. The integrated intensity was determined by fitting the rocking curves to a Lorentzian line-shape. When appropriate, $I_{\textbf{q}}$ was corrected by the cosine of the angle between the scan direction and \textbf{q} (the Lorentz factor)~\cite{Squ78}. In the work presented here, we measured the form factor for the first-order reflections which are all equivalent, so we may write it as $F(q)$. 

\section{Results and Discussion}

\subsection{Vortex Lattice Structure}

Previous VL studies on YBCO$_{7}$ \cite{Whi09, Whi11}, up to 10.8~T, showed first-order transitions from low field hexagonal structures to a high field rhombic phase. At high fields, the VL structure evolved  \textit{towards} a square configuration. Fig. 1(a) shows a typical (not quite square) diffraction pattern obtained at high field and base temperature. Fig. 1(b) shows the evolution of the VL shape as a function of field from 8 to 16.7~T for two different temperatures, and panel (c) shows the evolution with temperature on warming from a base of 2~K for two different fields. The evolution with field is seen to continue, passing \textit{through} a square configuration at around 12.5~T at base temperature.

London theory with non-locality (NL) corrections has had some success in describing the VL of superconductors with a fourfold axis~\cite{Kog97a, Kog97b} and in a conventional high-$\kappa$ material such as V$_3$Si~\cite{Yet99}. NL can give extra anisotropy about a crystal axis, in addition to the effective mass anisotropy in local theories. They can arise from two causes: either Fermi surface anisotropy, which was initially treated in combination with a constant superconducting energy gap $\Delta$~\cite{Kog97a, Kog97b}, and/or from an anisotropic $\Delta$~\cite{Kog97a, Fra97, Suz10}. In $d$-wave materials, the effective coherence length $\xi \propto   1/\Delta(k)$, such that non-local effects must always be important near nodes in the energy gap. However for fields up to 11~T, none of the NL theories agree with the observed structural evolution of the VL in YBCO$_{7}$, an observation which continues to hold for the present measurements \cite{Whi11}. We therefore prefer to rely on the predictions of microscopic calculations using Eilenberger theory~\cite{Ich99}. From these it is predicted that at high fields the nearest neighbor vortex directions will lie along the nodal directions of the order parameter. Our observations of the VL structural evolution suggest a field-driven change in the nodal positions, and we propose that superconducting states on the CuO chains may be responsible for this. 

The role of the chains in the superconductivity of YBCO$_{7}$ is illustrated by the value of the low field London penetration depth anisotropy $\sim 1.28$~\cite{Whi09, Whi11}, which implies an anisotropy in $n_{s} / m^{\ast}$ of $> 1.6$. As emphasized earlier, this value is higher than can be accounted for by plane superconductivity alone and the sign of the anisotropy implies an extra contribution to superfluid density, $n_s$, from the chains. Furthermore, it is clear  from the temperature-dependencies of $n_s$ that the superconducting order parameter has nodes on both the plane and chain Fermi surfaces~\cite{Whi14}. However, the  the quasi 1-D chain superconductivity appears to be less robust than its plane counterpart, since the $ab$ anisotropy reduces with field~\cite{Whi09, Whi11}. The behavior of the chain states in YBCO$_{7}$ is distinctly different from the chain states in YBa$_{2}$Cu$_{4}$O$_{8}$ \cite{Whi14}. That compound requires a multi-gap description, which corresponds to a proximity-effect single particle coupling between chain and plane states. This is in contrast to the single gap behavior in YBCO$_7$, which corresponds to a Josephson-like pair tunneling model for chain/plane coupling. This rules out a common description for chain-state behavior in Y-based cuprates. In the data presented here, we see that the anisotropy changes sign at high fields, since the VL structure passes \emph{through} square coordination. Since we are at a small fraction of $H_{c2}$ even at 17~T, we propose that this is an indication of a modification of the weaker chain superconductivity with field; in particular, the nodal positions on the chain Fermi surface may move. New ARPES experiments to confirm chain nodes at zero field, and high-field STM or detailed calculations may be able to confirm these ideas.

\begin{figure}[]
	\epsfig{file=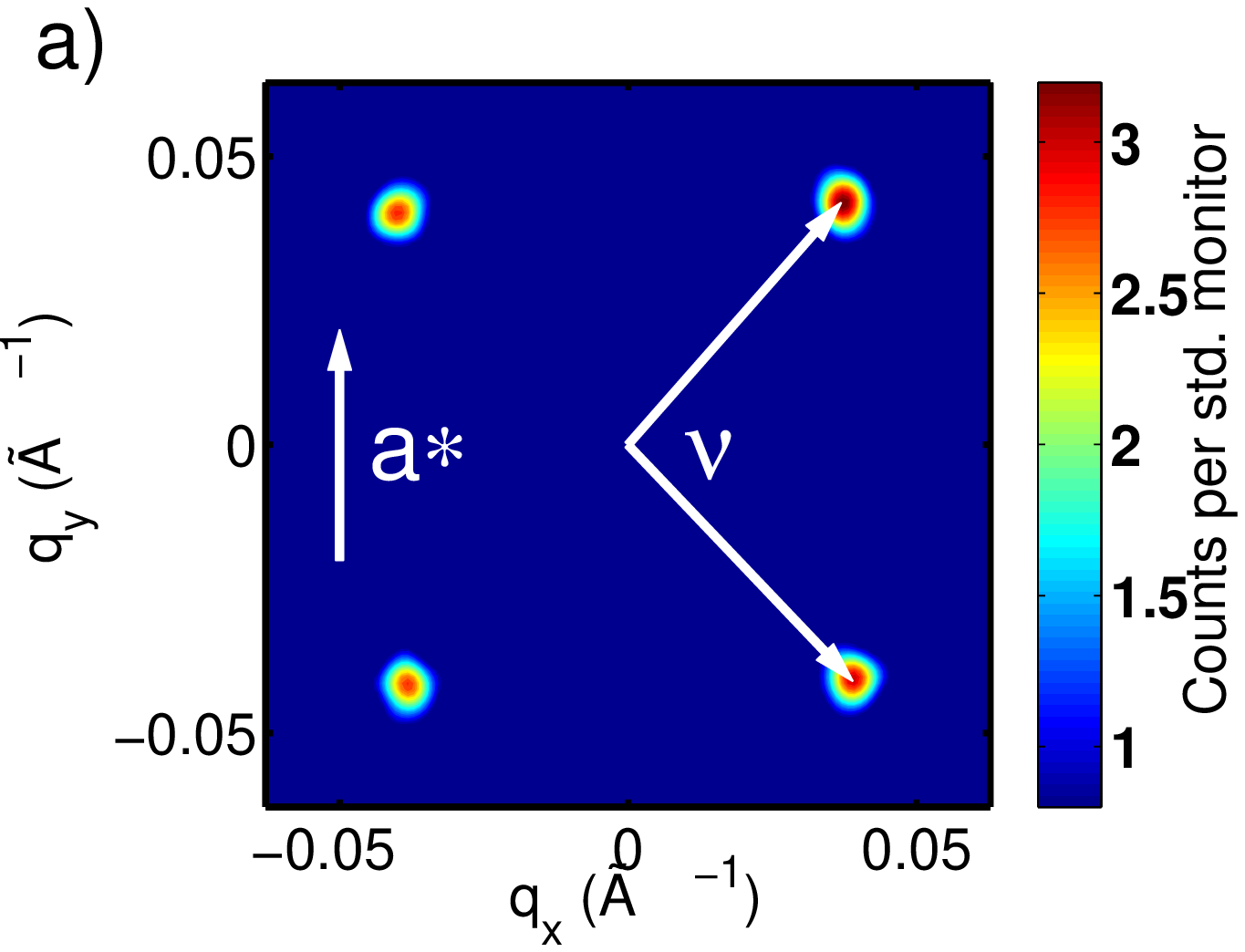, width=0.8\linewidth}
      \epsfig{file=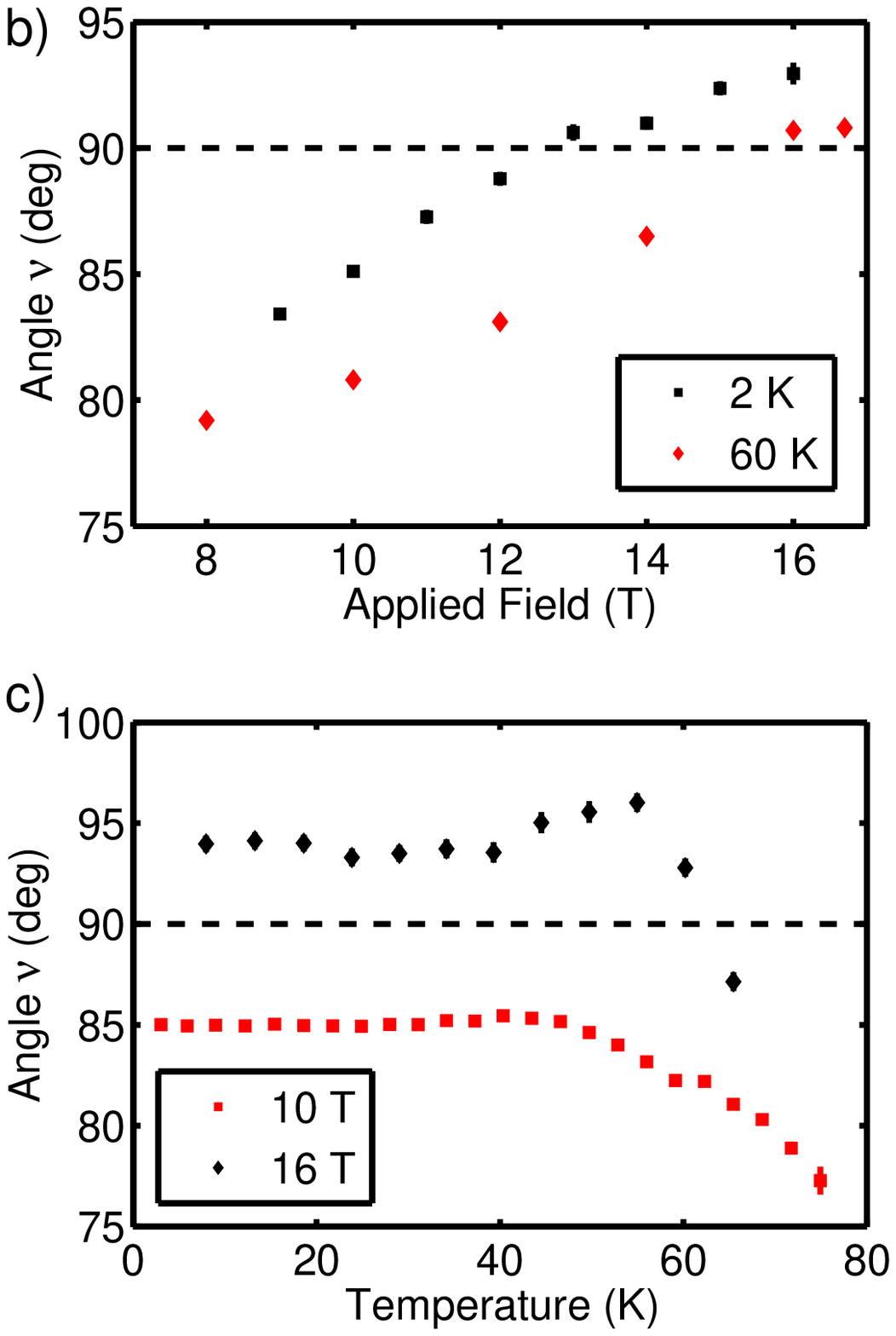, width=0.8\linewidth}
\caption{(Color online) (a) Typical diffraction pattern from the VL, taken in a field of 16 T at 2 K. The structure of the VL may be described by the opening angle $\nu$. (b) Graph showing the structural evolution of the VL as a function of applied field.  Measurements were taken after cooling while the field was oscillated slightly in value (OFC) to 60~K or 2~K, for fields in the range 8 to 16.7~T (c) Values of $\nu$ for a range of temperatures at applied fields of 10 and 16 T, taken on warming from base temperature after OFC.}
		\label{Graph1}
\end{figure}

%Include not locking in to field indep structure?

Fig.~1(c) shows the evolution of the VL structure with temperature. We first consider the behavior near $T_c$. It is predicted \cite{Kog97a, Fra97, Suz10} that NL will be reduced on approaching \textit{T}$_{c}$. This suppression would bring the VL structure towards the hexagonal structure expected by local electrodynamics, which is in general agreement with the results in figure 1(c).
The non-monotonic variation with temperature seen near 40-60~K is surprising; we believe this is intrinsic but since it occurs at a temperature where pinning effects are changing, we postpone discussion until these are considered.

%whilst we would expect this behaviour to be a result of annealing of the VL during warming through the irreversibility temperature (we recall that the data presented in figure \ref{Graph1}(b) were taken after preparing the lattice by OFC to base temperature, with measurements taken on warming), brief measurements also indicate that this `bump' also appears with measurements taken on \textit{cooling}.
%The non-monotonic variation with temperature seen near 40-60~K is at first surprising; we believe that this arises from annealing of the VL on warming through an irreversibility temperature. We recall that the data were taken by first OFC cooling to base, followed by warming to take measurements, with each measurement taking around 20-40 minutes depending of field.  We see that during warming the VL adjusts to a larger value of the angle $\nu$, which implies that in the absence of pinning the low-temperature value of $\nu$ would be even larger than we have observed. Other aspects of the data, considered later, give strong support for this pinning scenario.

\subsection{Variation of Form Factor with Field}

\begin{figure}[h]
	\epsfig{file=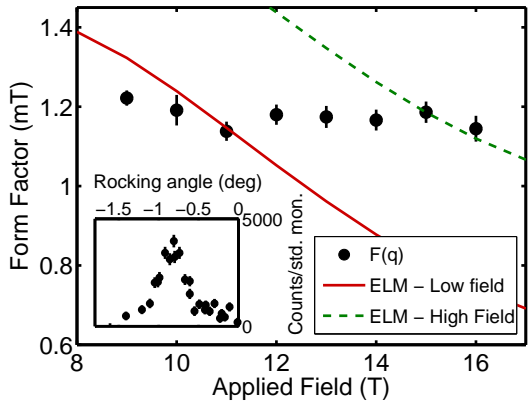, width=0.9\linewidth}
	\caption{(Color online) The variation with field of the VL form factor for the 1st order diffraction spots (shown in Fig.~\ref{Graph1}(c)) at a temperature of 2 K. Integrated intensities were obtained by fitting the data to a Lorentzian line-shape. The solid red line shows the extended London model (ELM) prediction (Eqn.~(2)), using parameter values obtained by fitting earlier data taken below 11 T~\cite{Whi11}, whilst the dotted green line shows a prediction, discussed in the text, from the ELM using smaller values of $\xi$.}
\label{Graph2}
\end{figure}
%Where did JSW get the model from? Can I cite them?
Now we turn to the spatial variation of magnetic field within the VL. To model the field dependence of the form factor, we have used the the local London model, extended to account for the \textit{ab} plane anisotropy of YBCO$_{7}$, and including a Gaussian cut-off term to account for the finite size of the vortex cores ~\cite{Whi11}:
\begin{equation}
F(\textbf{q}) = \frac{\langle B\rangle exp(-c(q_{x}^{2}\xi_{b}^{2} + q_{y}^{2}\xi_{a}^{2}))}{q_{x}^{2}\lambda_{a}^{2} + q_{y}^{2}\lambda_{b}^{2}}.
\label{LondonFF}
\end{equation}
Here, $\langle B \rangle$ is the average internal induction, $\lambda_{i}$ is the penetration depth arising from supercurrents flowing in direction $i$, $\xi_{i}$ is the coherence length along $i$, and $q_{x}$, $q_{y}$ are in-plane Cartesian components of the scattering vector, with $q_{x}$ parallel to \textbf{b}$^{\ast}$. The cut-off parameter \textit{c} accounts for the finite size of the vortex cores, and a suitable value for $c$ in our field and temperature range is 0.44~\cite{Whi11}.

Using the values for $\xi$ and $\lambda$ from previous experiments \cite{Whi11}, with $\lambda_{a} = 138 $~nm, $\lambda_{b} = 107$~nm, $\xi_{a} = 3.04$~nm and $\xi_{b} = 3.54$~nm, labeled as the `Low Field' model in Fig. \ref{Graph2}, we find that the form factor above 10 T is much larger than the extrapolation from low-field data. Taking smaller values of $\xi_{a} = 2.60$~nm and $\xi_{b} = 3.03$~nm with $\lambda$ unchanged provides the `High Field' line in Fig. \ref{Graph2}; however, this departs from the lower field data. Indeed, we find that no physically-reasonable constant values of the parameters fit the full field-range of our data. 

As noted earlier~\cite{Whi11}, the low-field fitted values for $\xi_{a}$ and $\xi_{b}$ are larger than expected and therefore not intrinsic, because the upper critical field estimated from them is too low for YBCO$_7$. These authors suggested that instead they reflect pinning effects. These remarks remain true for the present high-field results. The larger values of $\xi$ probably reflect frozen-in disorder in the VL giving a static Debye-Waller (DW) factor, reducing the scattered intensity. A mean square deviation $\langle u^{2}\rangle$ from straightness would contribute an additional term $exp(-q^{2}\langle u^{2}\rangle/4)$ to Eqn. (\ref{LondonFF}), which has the effect of increasing the Gaussian cut-off term and simulating a larger vortex core size. We deduce that the effects of disorder in our high-field range are relatively smaller, and that the actual values of $\xi$ are smaller than those obtained from fits.

We note that disorder in the VL leading to an over-estimation of the coherence length $\xi$ from SANS has been observed in other high $T_{c}$ materials. The superconductor La$_{2-x}$Sr$_{x}$CuO$_{4}$ (LSCO) gave distinctly different values for $\xi$ from SANS in comparison to heat capacity measurements \cite{Cha12, Wan08}. This was attributed to disorder in the VL leading to a similar over estimation of the coherence length to the situation we have here, although the disorder in LSCO was driven by electronic effects rather than the static pinning as is the case in close-to-optimally doped YBCO.

\subsection{Variation of the Vortex Lattice with Temperature}

\begin{figure}[h]
	\epsfig{file=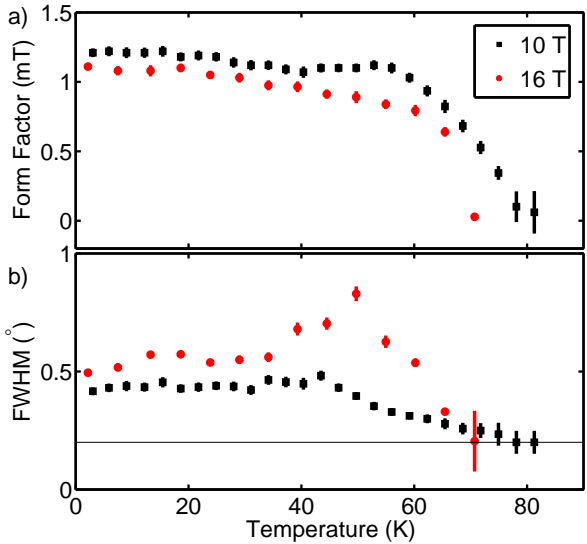, width=0.9\linewidth}
	\caption{(Color online) (a) The variation with temperature of the VL form factor at $10$ and $16$ T, with data taken on warming after OFC. The integrated intensity was calculated by fitting the rocking curves to a Lorentzian line-shape. (b) rocking curve widths vs. temperature, compared with the resolution width of the instrument (solid line).}
	\label{Graph3}
\end{figure}

Fig. 3 shows the variation of VL form factor and rocking curve width (FWHM) with temperature. The rocking curve width is the combined effect of both instrument resolution and the perfection of the VL~\cite{Cub92}, and is related to the correlation length, $\xi_{L}$, of the vortices along the field direction ~\cite{Yar94}. It can be seen that VL disorder dominates away from $T_{c}$. From such data, we can identify an `irreversibility temperature' $\sim$ 40-50~K below which the VL is `frozen in' and the rocking curve width and structure (Fig. 1(b)) do not change. Close to $T_{c}$, the FWHM of both data sets lie above the resolution limit, showing that pinning can \textit{still} distort the vortex lines away from straightness even though the VL as a whole can move. Near the irreversibility temperature, the 10~T and 16~T data exhibit differing behaviors, with the rocking curve width at 10~T decreasing towards $T_{c}$, whilst at 16~T the width \textit{increases}  before also narrowing on the approach to $T_{c}$. In the same region, the opening angle of the VL structure in fig. \ref{Graph1} has a peak. Further measurements, not reported here, show that this occurs both on warming and cooling, indicating that this is not due to annealing the VL on warming. We speculate that also in this case temperature/field dependent changes in the nodes, particularly in the chain order parameter, may be influencing the VL structure and pinning. 

The temperature dependence of $F(q)$ in Fig.~\ref{Graph3}, particularly at 10 T, may also represent such intrinsic effects, although we cannot rule out the effects of changes in a static DW factor. However, well below the irreversibility temperature both the VL structure and its lattice perfection remain fixed; hence the temperature dependence of $F(q)$ should reflect intrinsic properties of the sample. It has been suggested that the weak temperature dependence of $F(\textbf{q})$ seen at low temperatures is due to a stronger NL effects in large fields~\cite{Ami00, Whi11}. Following the non-local model employed earlier~\cite{Whi11} to fit data from YBCO at lower fields, we find that using the parameters for $\xi$ and $\lambda$ from Fig.~\ref{Graph2} provides a fit to the data at base temperature, and allowing these parameters to vary provides a good fit up to around 40 K. Above this region we expect that the static DW factor would reduce as the VL becomes able to move. This reduction in disorder leading to an increase in SANS intensity from a vortex lattice was first observed in BSCCO \cite{For96}, where thermal fluctuations at higher temperature was seen to smooth out the pinning potentials for the VL, leading to a re-emergence of the vortex lattice signal at higher temperatures. The effect of thermal fluctuations in YBCO, however, appears far less pronounced since the material is much more isotropic than BSCCO, and as such the vortex lines in YBCO do not decompose into vortex pancakes and the disorder is correspondingly much lower.

\subsection{Vortex Lattice Melting}

\begin{figure}[h]
	\epsfig{file=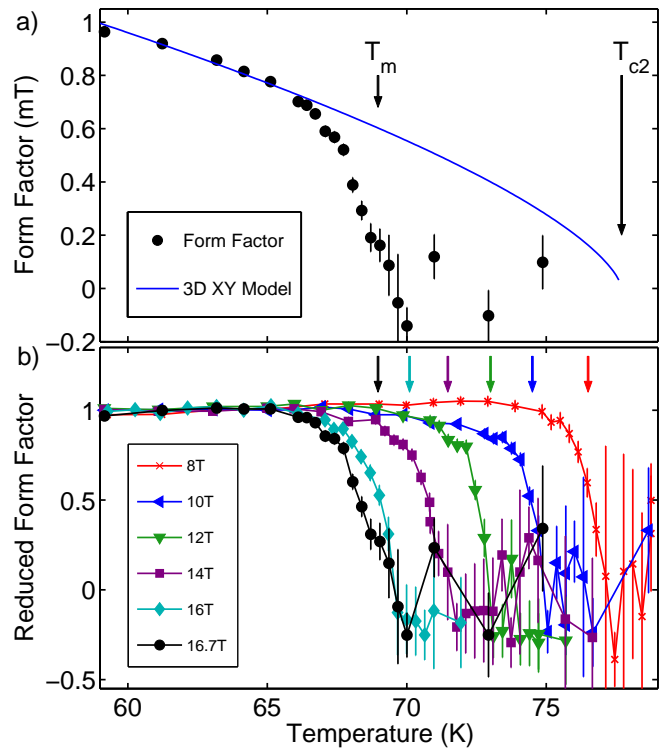, width=1\linewidth}
	\caption{(Color online) Graph showing the variation with temperature of $F(q)$ across the VL melting transition. (a) The raw form factor is shown for an applied field of $16.7$ T, alongside the predicted variation from the $3$D-XY model. (b) The reduced form factor is shown for fields between $8$ and $16.7$ T, with the reduced form factor obtained by dividing the data by the prediction from the $3$D-XY model. The arrows indicate melting temperatures, obtained from a linear fit versus temperature of the fall to zero of the intensity. The noise above the melting temperature appears larger both because of the square root relating intensity to form factor, and also because the plotted quantity is divided by the function in Eqn. 2, which is tending to zero.}
	\label{graph4}
\end{figure}

In Fig. 3, the form factor at 16~T shows a sudden drop close to 70~K. A more detailed scan between 60 and 80~K at 16.7~T is shown in Fig.~4(a), together with a solid line giving the superfluid density from the 3D-XY model~\cite{Kam94}. This model can be used to give the VL form factor as:
\begin{equation}
F  \propto \big( 1-T/T_{c2}(B) \big) ^{n},
\end{equation}
with $T_{c2}(B)$  taken from work by Junod \textit{et al} \cite{Junod97}, and $n=0.66$. It can be seen in Fig.~\ref{graph4}(a) that the form factor data drops well below the model prediction. The falloff will begin as a (true) Debye-Waller effect arising from the increasing thermally-induced excursions of the VL from equilibrium, followed by VL melting when the displacements become a large enough fraction $c_{L}$ of the the vortex line spacing~\cite{Kierfeld04}. Here, $c_{L}$ is called the Lindemann number and is expected to lie in the range 0.15 to 0.3. We can estimate the reduction of intensity due to the DW factor, $exp(-q^{2}\langle u^{2}\rangle/2)$, using  $\langle u ^{2}\rangle = c_{L}^{2}a^{2}$, where $a$ is the spacing between vortex lines. With $c_{L} = 0.2$, we find that before melting the intensity of a first-order diffracted spot would already have fallen to a fraction $\sim0.45$ of that for a perfect lattice. In Fig.~4(b) are shown the data for a range of fields plotted as a fraction of the 3D-XY variation. The melting line \textit{B}$_{m}$(\textit{T}) derived from these results is very close to that found by heat capacity measurements \cite{Rou98} of the lattice-liquid transition. However, in addition, the present data give microscopic information about the VL structure as melting occurs. Macroscopic measurements have shown that a transition point between first and second order melting is dependent on pinning disorder and oxygen content \cite{Kha07, Nis00, Rou98}. First-order melting is expected in our case, because macroscopic measurements on YBCO$_{7}$ from the same source as our sample show that the first order transition continues up to $\sim 30$ T~\cite{Shi02}. We hesitate to compare our SANS observations of VL melting in YBCO$_7$ with other observations using the same technique, because all other measurements were on very different materials. The high $T_c$ superconductor BiSCCO is a \emph{very} anisotropic material, so that the vortex lines are very flexible and easily disrupted by pinning or thermal fluctuations; VL melting occurs at far lower fields than in YBCO, although there are some interesting suggestions of DW effects in the data~\cite{Cub93,For96}. An earlier paper on YBCO~\cite{Aeg98} reported results in a much lower field range on a heavily twinned sample, while observations on an ultra-pure sample of niobium using both SANS and high-resolution heat capacity techniques~\cite{Bow10} surprisingly show no sign of analogous VL melting in this non-layered low $T_c$ material.

The intensity of Bragg reflections is predicted to drop sharply to near zero upon melting of the VL due to the loss of crystalline order, which would smear the spots into a ring, and we estimate this would reduce the signal at the reflection by an order of magnitude. However, even after counting for an extended period, no signal from the vortex liquid was detected above the transition. Should the rocking curve width of the vortex liquid remain similar to that of the vortex solid - and we note that the FWHM of the VL remained narrow during the melting transition - then we can estimate the Lindemann number required for a DW factor to reduce the scattering from the vortex liquid to the point where we could no longer observe it. We find $c_{L}$ to be at least $\sim$0.25, which is within the expected range of values~\cite{Kierfeld04}. Hence, the absence of a SANS signal from the vortex liquid does \emph{not} imply the non-existence of this phase. Whilst we expect a first order transition to be sharp, we see that these data do not appear to have a sharp fall to zero at the melting transition, although this may be obscured by the initial fall due to the Debye-Waller factor. We note that this spread also reflects the slight spread in $T_{c}$ of our sample.
~\\

\section{Conclusions}

In conclusion, we report in this paper the microscopic investigation of the structure of vortex matter in a high-$T_c$ superconductor in a new high-field range. We have observed the structural evolution of the VL rhombic phase in YBCO$_{7}$, showing no field independent structure and suggesting a field-driven change in the superconducting states of the CuO chain layers. The evolution of the VL form factor with field away from the low-field London+core model indicates that the VL will remain measurable by SANS well above our maximum field, allowing for measurements to be extended further when equipment permits. The deviation of both the field and temperature dependence of the VL form factor from the expected behavior in the London model, combined with the unusual variation in VL structure with increasing temperature, suggests that a static Debye-Waller factor is present in the low field and temperature data. This effect is reduced both at high field and temperature, and confirms that estimates of the coherence length obtained from lower field data are too long \cite{Whi11}. Observations of the VL at high fields and high temperatures show the effects of thermal displacements on the VL perfection prior to the VL melting transition, and provide different information from macroscopic measurements. Our measurements are consistent with a first order lattice to liquid transition, with the intensity scattered from the vortex lines falling sharply across the transition to an unmeasurably low value in the vortex liquid state. In summary, our study provides a further perspective on vortex matter under extreme conditions.

\section{Acknowledgements}

We acknowledge funding from the UK EPSRC, the University of Birmingham, the Swiss NCCR and its program MaNEP; also support from the ILL, where the measurements were performed and the assistance of S. M\"uhlbauer and A. Heinemann at SANS-I of FRM-II with additional measurements not reported here.

\bibliographystyle{apsrev4-1.bst}
\bibliography{YBCO_2014_V5}

\end{document}